\documentclass[journal,twocolumn]{IEEEtran}
\usepackage{amsmath,bm}
\usepackage{amssymb}
\usepackage{latexsym}
\usepackage{multicol}
\usepackage{amsfonts}
\usepackage{graphicx}
\usepackage{subfig}
\usepackage{float}
\usepackage{color}
\usepackage{cite}
\usepackage{multirow}
\usepackage{diagbox}
\usepackage{algorithm}
\usepackage{algorithmic}
\usepackage{booktabs}
\begin{document}

\title{Target Localization with Jammer Removal Using Frequency Diverse Array}
\author{Qi Liu,
        ~Jingwei Xu,~\IEEEmembership{Member, IEEE},
        ~Zhi Ding,~\IEEEmembership{Fellow, IEEE},
        and~Hing Cheung So,~\IEEEmembership{Fellow, IEEE}
%
\thanks{Q. Liu is with the Department of Electrical and Computer Engineering, National University of Singapore, Singapore (E-mail: qiliu47-c@my.cityu.edu.hk).}
\thanks{J. Xu is with the National Laboratory of Radar Signal Processing, Xidian University, Xi'an, Shaanxi, 710071, China (E-mail: xujingwei1987@163.com).}
\thanks{Z. Ding is with the Department of Electrical and Computer Engineering, University of California at Davis, Davis, CA 95616 USA (E-mail: zding@ucdavis.edu).}
\thanks{H.C. So is with the Department of Electrical Engineering, City University of Hong Kong, Hong Kong, China (E-mail: hcso@ee.cityu.edu.hk).}
}
\maketitle

\begin{abstract}
A foremost task in frequency diverse array multiple-input multiple-output (FDA-MIMO) radar is to efficiently obtain the target signal in the presence of interferences. In this paper, we employ a novel "low-rank + low-rank + sparse" decomposition model to extract the low-rank desired signal and suppress the jamming signals from both barrage and burst jammers. In the literature, the barrage jamming signals, which are intentionally interfered by enemy jammer radar, are usually assumed Gaussian distributed. However, such assumption is oversimplified to hold in practice as the interferences often exhibit non-Gaussian properties. Those non-Gaussian jamming signals, known as impulsive noise or burst jamming, are involuntarily deviated from friendly radar or other working radio equipment including amplifier saturation and sensor failures, thunderstorms and man-made noise. The estimation performance of the existing estimators, relied crucially on the Gaussian noise assumption, may degrade substantially since the probability density function (PDF) of burst jamming has heavier tails that exceed a few standard deviations than the Gaussian distribution. To capture a more general signal model with burst jamming in practice, both barrage jamming and burst jamming are included and a two-step "Go Decomposition" (GoDec) method via alternating minimization is devised for such mixed jamming signal model, where the $a$ $priori$ rank information is exploited to suppress two kinds of jammers and estimate the desired target.  Simulation results verify the robust performance of the devised scheme.

\end{abstract}

\begin{IEEEkeywords}
Target localization, Multiple-input multiple-output (MIMO), Frequency diverse array (FDA), Low-rank matrix approximation, Mixed jamming signals.
\end{IEEEkeywords}

\IEEEpeerreviewmaketitle

\section{INTRODUCTION}

\IEEEPARstart{T}{arget} localization with multiple-input multiple-out (MIMO) radar system has attracted much attention in recent years \cite{ColocatedMIMO,8477186,Qi2015,8280507}. The context of this work belongs to the co-located MIMO radar, where the MIMO technique can greatly enhance the angular converge of radar system by deverging orthogonal waveforms. Like conventional co-located MIMO radar, the array manifold caused by time delay only depends on the angle, unable to intractably suppress the interferences which share the same angle but different range with targets.

The concept of frequency diverse array (FDA) was first proposed in \cite{FDA}, where it is based on the employment of a small frequency increment across the array elements. It is worth noticing that relative to the carrier frequency and bandwidth of the transmitted waveform, the frequency increment is small and negligible. In general, FDA induces additional range-time-dependent phase information in the transmit waveforms, resulting in a change of beampattern as a function of the range, angle and time. Angle-range-dependent beampattern provides many promising applications, such as synthetic aperture radar (SAR) and moving target indication, space-time adaptive processing for clutter suppression and so on \cite{FDA-STAP2015, FDA-SAR2018}. In general,  the advantage of mitigating range ambiguous clutter with only a single pulse repetition frequency makes FDA suitable for target detection of radar mounted on the super-high speed space vehicle.

\subsection{Prior Work}
By exploiting additionally controllable degrees of freedom (DOFs), range ambiguous clutter is effectively eliminated by FDA-STAP radar \cite{FDA-STAP}, which also enjoys significant performance improvement over traditional phased-array radar. To further exploit the DOFs in range and angle domains, FDA is merged with MIMO radar in \cite{FDA-MIMO2009, FDA-MIMO2013}, where the frequency offset is much smaller than the modulation bandwidth. Although additional DOFs of FDA have been studied, the angle-Doppler-defocusing of a fast-moving target is rarely noticed, which causes mismatch between the presumed and true target of steering vectors. As a result, the performance of the traditional minimum variance distortionless response beamformer degrades dramatically. To cope with the mismatch, robust adaptive beamforming (RAB) methods have been suggested \cite{1420809, 1166614, 1206680}. In the FDA-STAP radar, the target signal non-ideally focuses in angle and Doppler domains, which requires a large feasible region to retrieve the true steering vector. However, large feasible set, e.g., spherical set, may contain the zero point and generates trivial solutions. Hence, a new RAB approach is devised by properly designing the feasible region \cite{RAB-FDA-STAP}. In \cite{Phased-MIMO2013}, the FDA framework is adopted in the phased-MIMO radar by dividing FDA into subarrays and transmitting the coherent waveform to maintain range-angle-dependent beampattern. In \cite{FDA2015}, an unambiguous joint range-and-angle estimation method is proposed, where the adaptive weight in the joint transmit-receive domains is constructed by using the $a$ $priori$ of noise and interference variance information to estimate the target parameters. Moreover, a range ambiguity resolution technique is derived through a proper  frequency increment. In \cite{Nonuniform-FDA}, A nonuniform FDA radar is utilized for target imaging in range and angle domains, where the improved range resolution is achieved by enlarging the frequency increment. However, both of these parameter estimates are based only on barrage jamming signal, which may oversimplify the practical scenarios.

\subsection{Contribution}
In studying the joint range and angle estimation of FDA-MIMO radar, their estimation in the presence of burst jamming signal is rarely considered, which can significantly degrade the accuracy of the existing methods. In this work, targeting the more practical mixed jamming signals, we devise a novel joint range and angle estimator for FDA-MIMO radar. Different from robust principal component analysis (PCA), the proposed scheme is based on a "low-rank + low-rank + sparse" decomposition, where a Frobenius norm minimization with $a$ $priori$ rank is added to suppress false targets resulting from the mixed jamming signals in "low-rank + sparse" decomposition algorithms, e.g., the GoDec method \cite{GoDec}. The main contributions of this paper are summarized as follows:

1) To the best of our knowledge, this is the first time that the burst jamming signal is exploited in FDA-MIMO radar for joint range and angle estimation, which exhibits non-Gaussian property.

2) We propose to employ the "low-rank + low-rank + sparse" decomposition model instead of robust PCA, which is more explicit to understand for the proposed problem; To successfully separate low-rank desired matrix, a sampling scheme is added into the procedure of down converting, matched filtering and storing, where the reasonability of model is analyzed.

3)  A two-step GoDec method with alternating minimization to avoid false targets in the existing GoDec method, which is parameter-free and easily implemented.

\subsection{Organization}

The rest of this paper is organized as follows. We present the signal and noise models with interferences of co-located FDA-MIMO radar in Section II. In Section III, we study a low-rank matrix approximation model and propose a robust joint range and angle estimator for FDA-MIMO radar based on the alternating minimization, where the range and angle parameters of targets are obtained from the two-dimensional (2-D) Fourier transform. Simulation results are provided in Section IV to show the effectiveness of the proposed method. Finally, conclusions are drawn in Section V.

\section{PROBLEM FORMULATION AND PRELIMINARIES}

\subsection{Signal Model of Colocated FDA-MIMO Radar}
A standard co-located FDA-MIMO radar system consisting of $M$-element uniform linear transmit array and $N$-element uniform linear receive array, is considered throughout this paper  \cite{FDA2015}. Assume that the total transmit energy is $E$ and each element has the same energy $E\over M$. Since the transmit array is an FDA, orthogonal waveforms are emitted  with minimum frequency increments along the array elements. Under the assumption of orthogonality, the transmit waveforms can be effectively separated in the receiver. To be specific, the narrow-band complex signal transmitted by the $m$th element is modeled as:
\begin{align}
s_m(t) = \sqrt{E\over M} \phi_m(t)e^{j2\pi f_mt}, ~0 \le t \le T_\iota, ~m = 1, ..., M
\end{align}
where $T_\iota$ is the radar pulse duration, the carrier frequency of FDA is $f_m = f_0 + (m-1)\Delta f$, $f_0 = f_1$ is the reference carrier frequency and $\Delta f$ is the frequency increment across the array elements. Compared with the carrier frequency and bandwidth of the transmit waveforms, the frequency increment is relatively small and can be negligible. Let $\phi_m(t)$ be the transmitted baseband complex waveform, satisfying the unit-energy and orthogonality properties, i.e., $\int_0^{T_\iota} \phi_m(t)\phi_m^*(t)\, dt = 1 $ and
\begin{align}
\int_0^{T_\iota} \phi_n(t)\phi_m^*(t - \tau_m)e^{j2\pi\Delta f(m-n)t}\, dt = \begin{cases} 0, m \neq n, \forall \tau \\
1, m = n, \tau =0, \end{cases} \notag
\end{align}
respectively, where $\tau_m$ is the time shift and the superscript $^*$ represents the conjugate operator. Note that the orthogonality condition of FDA-MIMO radar is more feasible than that of the standard MIMO radar, that is, $\int_0^{T_\iota} \phi_n(t)\phi_m^*(t - \tau)\, dt = 0, m \neq n, \forall \tau$. It is because  when $\Delta f = 1/T_\iota$, the orthogonality condition of FDA-MIMO radar can be achieved by any waveforms with constant modulus.

Let $\nu$ be the radial velocity of the target and the wavelength of $m$th transmit antenna is $\lambda_m = c/{f_m}$ with $c$ being the speed of light, then the Doppler frequency is expressed as $f_{d,m} = {2\nu/ \lambda_m}$, which can be approximated by $f_{d} = {2\nu/ \lambda_0}$ with the reference wavelength $\lambda_0$. For a monostatic MIMO radar, given an arbitrary far-field target of interest with angle-range pair $(\theta, r)$, the echoed signal of the $m$th element is expressed as:
\begin{align}
\label{xn-t}
x_n(t) = \sum_{m = 1}^{M}\sqrt{E\over M}\xi \phi_m(t-\tau_{m,T}-\tau_{n,R})\notag\\
\times e^{j2\pi (f_m + f_{d})(t-\tau_{m,T}-\tau_{n,R})},
\end{align}
where $\xi$ is the complex-valued reflection coefficient of the target, $\tau_{m,T} = {r_{m,T} \over c}$ and $\tau_{n,R} = {r_{n,R} \over c}$ are the transmit and receive time delays from the $m$th element to the target, respectively. Taking the first element as the reference, the range differences between $m$th element and the reference element are approximated by
\begin{align}
\label{r-mn}
r_{m,T} &\approx r - (m-1)d_T\sin(\theta) \notag\\
r_{n,R} &\approx r - (n-1)d_R\sin(\theta)
\end{align}
where $r$ is the range of target for the reference element, $d_T$ and $d_R$ denote the interspacings of transmit and receive antennas, respectively.

Substituting \eqref{r-mn} into \eqref{xn-t} yields (See Appendix):
\begin{align}
x_n(t) &\approx\sum_{m = 1}^{M}\sqrt{E\over M}\mathcal{R}_{f}\phi_m(t-\tau_{m,T}-\tau_{n,R}) e^{j2\pi (m-1)\Delta ft} \notag\\ &\times {e^{{-j2\pi }({{(m-1)\Delta f2r \over c}}-{(m-1)d_T f_0\sin\theta\over c}-{(n-1)d_Rf_0\sin\theta\over c})}}
\label{Rf}
\end{align}
where $\mathcal{R}_{f} =\xi{e^{j2\pi f_0(t-{2r\over c})}}{e^{j2\pi f_d(t-{2r\over c})}}$ is the merged reflection coefficient. The first term is important because it makes the array radiation pattern depend on the range and frequency increment. The second term is simply for the conventional phased-array factor seen frequently in array signal processing theory. For the simplicity of notation, we define:
\begin{align}
\mathbf a_t(t) &= [1, e^{j2\pi\Delta ft}, ..., e^{j2\pi(M-1)\Delta ft}]^T \notag\\
\mathbf a_T(\theta) &= [1, e^{j2\pi{d_Tf_0\sin \theta\over c}}, ..., e^{j2\pi(M-1){d_Tf_0\sin \theta\over c}}]^T \notag\\
\mathbf a_r(r) &= [1, e^{-j4\pi{\Delta fr\over c}}, ..., e^{-j4\pi(N-1){\Delta fr\over c}}]^T.
\end{align}
where the superscript $^T$ denotes the transpose operator.

Hence, in the single-antenna structure, the returned signal in \eqref{Rf} at the $n$th element is concisely expressed as:
\begin{align}
x_n(t) &=\sum_{m = 1}^{M}\sqrt{E\over M}\mathcal{R}_{f}\phi_m(t-\tau_{m,T}-\tau_{n,R})\notag\\
&~~~~ \times \mathbf a_t(t)\odot\mathbf a_r(r)\odot\mathbf a_T(\theta) \times e^{-j2\pi(n-1){d_Rf_0\sin\theta\over c}}.
\label{xodot}
\end{align}
where $\odot$ denotes the Hadamard product.

The received signal in \eqref{xodot} is down converted and fed into a bank of matched filters $h_m(t) = \phi_m(t)e^{j2\pi(m-1)\Delta ft}$ to separate the transmit signals. Since $x_{nm}$ reaches its normalized amplitude peak 1 at $t = \tau_{m,T}-\tau_{n,R}$, at the output of the $m$th matched filter of the $n$th antenna, the filter output $x_{nm}$ at $t = \tau_{m,T}-\tau_{n,R}$ is written as:
\begin{align}
x_{nm} &= \int_0^{T_l}x_n(\mu)\phi_m^*(\mu - t)e^{j2\pi\Delta f(m-n)\mu}\, d\mu \notag\\
&= \sqrt{E\over M}\mathcal{R}_{f} \mathbf a_r(r)\odot\mathbf a_T(\theta) \times e^{-j2\pi(n-1){d_Rf_0\sin\theta\over c}}.
\end{align}

Therefore, the synthesized output of target signal can be expressed as a concise form:
\begin{align}
\mathbf{x_s} = \sqrt{E\over M}\mathcal{R}_{f}  \mathbf a_R(\theta)\otimes\mathbf a(r,\theta)
\label{y}
\end{align}
where $\otimes$ denotes the Kronecker product operator, the new transmit steering vector $\mathbf a(r,\theta) = \mathbf a_r(r)\odot\mathbf a_T(\theta)$ and receive steering vector
\begin{align}
\mathbf a_R(\theta)= [1, e^{j2\pi{d_Rf_0\sin \theta\over c}}, ..., e^{j2\pi(N-1){d_Rf_0\sin \theta\over c}}]^T \notag.
\end{align}
In this case, \eqref{y} is similar to a phased-array MIMO radar output except $\mathbf a_r(r)$ \cite{Liu2018}. Especially, for $\Delta f = 0$, $\mathbf a(r,\theta) = \mathbf a_T(\theta)$ simplifies to the standard phased-array, which reveals that the phased-array can be regarded as a special FDA with $\Delta f = 0$. It is worth noticing that different from the conventional uniform linear array (ULA) phased-array MIMO radar, the transmit steering vector of FDA-MIMO is range-angle-dependent, which leads to the fact that targets may be arbitrarily distributed in the FDA-MIMO radar spectrum, while targets of the MIMO radar are diagonally distributed in the transmitter-receiver spectrum distribution. Hence, it is of great importance since it provides local maxima at different range cells and can be used to suppress range-dependent interferences and clutter. 

\subsection{Jamming and Noise Model}
We assume that $\mathbf{x_n}(t)$ represents the Gaussian noise component with zero-mean, which is white in both spatial and temporal domains. Then, denoting $\sigma_n^2$ as the noise power for a single channel and a single pulse, the covariance matrix of the output noise component is:
\begin{align}
\mathbf {R_n} = \mathbb{E}\{\mathbf{x_n}(t)\mathbf{x_n}^H(t)\}=\sigma_n^2\mathbf{I}_{MN}
\end{align}
where $^H$ denotes the conjugate transpose operator, $\mathbb{E}\{\cdot\}$ is the expectation operator and $\mathbf I_{MN}$ is the $MN \times MN$  identity matrix.

Barrage or noise jamming, whether an electronic countermeasure or accidental, has the potential of deteriorating the detection performance of a radar by superimposing Gaussian noise in the receiver bandwidth \cite{Ahmed2014}.  However, non-Gaussian noise such as impulsive noise exists in real-life, crossing various frequency bands. This kind of noise can be caused by the internal circuit such as amplifier saturation, sensor failures, or from the external interference like thunderstorms and man-made noise. More importantly, it may be caused by the friendly operating devices. Therefore, except for the effect of output noise, the interferences of radar with accurate target localization technique in radar system \cite{RadarSystem} comes from two-folds: the jamming signals from the jamming devices of enemy radars (denoted as $\bf x_i$) and the radio equipments of friendly radars (denoted as $\bf x_e$), which are referred to as barrage and burst jammings, respectively. Next, the noise models of mixed jamming signals are built as follows.

\begin{figure}[!htbp]
\begin{center}
\includegraphics[width=8.5cm]{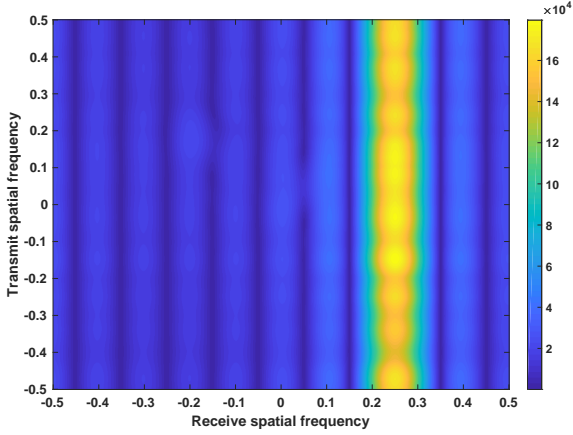}
\end{center}
\caption{Spectrum of barrage jamming in frequency domain}
\label{Jamming}
\end{figure}

The barrage jamming from enemy radars denotes all noises input into the mixers including background noise and thermal noise of frontier circuits, and circuit noise introduced by mixer and matched filter in each channel \cite{RGui2018}.  All jammers perform independently and they transmit independent envelops. Suppose $L$ jamming signals from jamming devices imping on the array, the receive interference of jamming signals corresponding to directions $\theta_j, j = 1, ..., L$ is constructed by \cite{jamming}
\begin{align}
\mathbf {x_i} = \sum_{j=1}^L \gamma_j \mathbf a_R(\theta_j) \otimes \mathbf n_j \label{Barrage}
\end{align}
where both $\gamma_j \sim \mathcal{N}(0, \sigma_j^2)$ and $\mathbf n_j \sim \mathcal{N}(0,1)$ are Gaussian distributed. The spectrum of jamming signal $\bf x_i$ is plotted in Fig. \ref{Jamming}, where the interference-to-noise ratio (INR) equals 30 dB and the jamming signal is impinging from the direction of $30^\circ$. As only one jamming signal is considered, the spectrum is focused in the receive spatial frequency domain. In contrast, the spectrum is uniformly-distributed in the transmit spatial frequency domain. This is consistent with the noise-like steering vector in \eqref{Barrage}. Due to the fact that rank$(\mathbf A\otimes \mathbf B)$ = rank$(\mathbf A)$rank$(\mathbf B)$, rank$( \mathbf a_R(\theta_j) \otimes \mathbf n_j)$ = $LM$, where $ \mathbf a_R(\theta_j)  \in \mathbb{C}^{N\times L}$ and $\mathbf n_j \in \mathbb{C}^{M\times T_\ell}$, $L < M, N < T_\ell$.  In this work, only one jamming signal is exploited, it turns out that the rank of $\bf x_i$ is $M$. 

\begin{figure}[!htbp]
\begin{center}
\includegraphics[width=8.5cm]{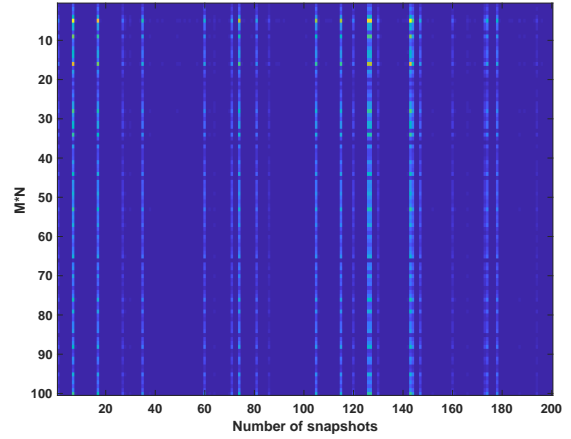}
\end{center}
\caption{Distribution of burst jamming in time domain}
\label{InterferenceS}
\end{figure}

On the other hand, since the jamming signal $\bf x_e$ from nearby operational radar is like thunderstorms, which behaves as burst noise or impulsive noise, it exhibits non-Gaussian property, and is frequently encountered in many radio systems, radar imaging and electronic reconnaissance in military \cite{AMZ2012, MIT2001, PCE2003,YH2019}. Herein, Gaussian mixture model (GMM) \cite{GMM,Qi2019} is adopted to model $\bf x_e$. The probability density function of the two-term Gaussian mixture noise is constructed by
\begin{equation}
p_n(n)=\sum_{i=1}^{2}{c_i\over{\pi\sigma_i^2}}\exp\left({- {{|{n}|^2}\over{\sigma_i^2}}}\right)
\end{equation}
where $0 \le c_i \le 1$ and $\sigma_i^2$ are the probability and variance of $i$th term, respectively, with $c_1+c_2=1$. If $\sigma_2^2 \gg \sigma_1^2$ and $c_2<c_1$ are selected, large noise samples of variance $\sigma_2^2$ occurring with a smaller probability $c_2$ can be viewed as impulsive noise embedded in Gaussian background noise of variance $\sigma_1^2$. Therefore, GMM can well model the phenomenon in the presence of both Gaussian and impulsive noises. In our simulations, $\sigma_2^2 = 100 \sigma_1^2$, $c_2 = 0.1$. Hence, there are $10\%$ noise samples that are considered as impulsive noise. The spectrum of the burst noise data $\bf X_e$ is shown in Fig. \ref{InterferenceS} ($\bf X_e$ is a matrix of $\bf x_e$ when collecting all snapshots), where forms multiple strong "stripes" in range-and-angle time domain. It is obvious that the burst noise has sparsity crossing different snapshots, which makes sense because these sensor outputs corrupted by the jammers at the $t$th snapshot will turn to large values when the jammer signals arrive at some sensors at time $t$, so that the element modulus of the $t$th column of $\bf X_e$ will be large with a high probability.

In the presence of noise, the synthesized output of signal plus total noise can be represented as:
\begin{align}
\mathbf y &= \sqrt{E\over M}\mathcal{R}_{f}  \mathbf a_R(\theta)\otimes\mathbf a(r,\theta) + \sum_{j=1}^L \gamma_j \mathbf a_R(\theta_j) \otimes \mathbf n_j + \mathbf {x_e} + \mathbf {x_n} \notag\\
&= \mathbf {x_s} + \mathbf{x_i} + \mathbf {x_e} + \mathbf{x_n}
\end{align}

In the total data snapshots, the signal model can be formulated into 2-D time domain as a matrix form:
\begin{align}
\mathbf Y = \mathbf {X_s} + \mathbf {X_i} + \mathbf {X_e} + \mathbf {X_n},
\label{matrixmodel}
\end{align}
where $\mathbf Y \in \mathbb{C}^{MN \times T_l}$. Let $\mathbf d(t) = [\rho_1(t), \cdots, \rho_K(t)]^T \in \mathbb{C}^{K\times 1}$ contain the targets scattering coefficients vector for the $K$ emitting signals, where $\rho_q(t) =  \sqrt{E\over M}\mathcal{R}_{f} , q = 1, \cdots, K$ are complex scattering coefficients of $K$ targets. Hence, for $K$ targets, the synthesized output of signal is $\mathbf {x_s}= \mathbf A_{R,T}\mathbf d(t)$, where the steering matrix $ \mathbf A_{R,T} \in \mathbb{C}^{MN\times K}$ is composed of steering vectors $ \mathbf a_R(\theta_q)\otimes\mathbf a(r_q,\theta_q), q = 1, \cdots, K$. It is easy to find rank($\mathbf {X_s}$) $= K$ in the total data snapshots due to the fact rank($\mathbf A_{R,T}\mathbf D$) $\le$ min(rank($\mathbf A_{R,T}$) = $K$, rank($\mathbf D$) = $K$). $\mathbf D$ is a matrix of $\mathbf d(t) $ when collecting all snapshots.  The advantage of this 2-D time model in \eqref{matrixmodel} is that the signal matrix has low-rank property, which is the basis of the proposed method. Different with \cite{FDA2015}, our work also considers the jamming signal from the burst noise performing as non-Gaussian property, which reveals that the signal model in \cite{FDA2015} can be regarded as a special case of our work without $\mathbf {X_e}$. In \cite{FDA2015}, due to the $a$ $priori$ information of total noise variance, the interference-plus-noise covariance matrix $\mathbf Q = \mathbf R + \mathbf{R_n}$ in receive domain is utilized to construct the adaptive weight to eliminate the interference and Gaussian noise, where $\mathbf R$ is the covariance matrix of $\bf X_i$. However, it is obviously not practical. Compared with \cite{FDA2015}, as for the unknown burst noise from working radio equipments, it is intractable to build an adaptive weight to suppress the noise and total interferences. Thus, since the range and angle of target are coupled in transmit steering vector, they cannot be directly estimated from \eqref{matrixmodel}. Note that our work deals with the joint range and angle estimation based on the original data matrix $\mathbf Y \in \mathbb{C}^{MN \times T_l}$, differing from \cite{FDA2015} with the reconstructing compensated data $\mathbf Y_{\rm comp} \in \mathbb{C}^{M \times NT_l}$.

\section{Robust joint range-and-angle estimator}
To derive a joint range and angle estimation method in the case of mixed friendly and hostile radar interferences, low-rank matrix approximation model is introduced in our work.

\subsection{Low-Rank Matrix Approximation Model}

\begin{figure}[!htbp]
\centering
\begin{minipage}{4cm}
\includegraphics[width=4.5cm]{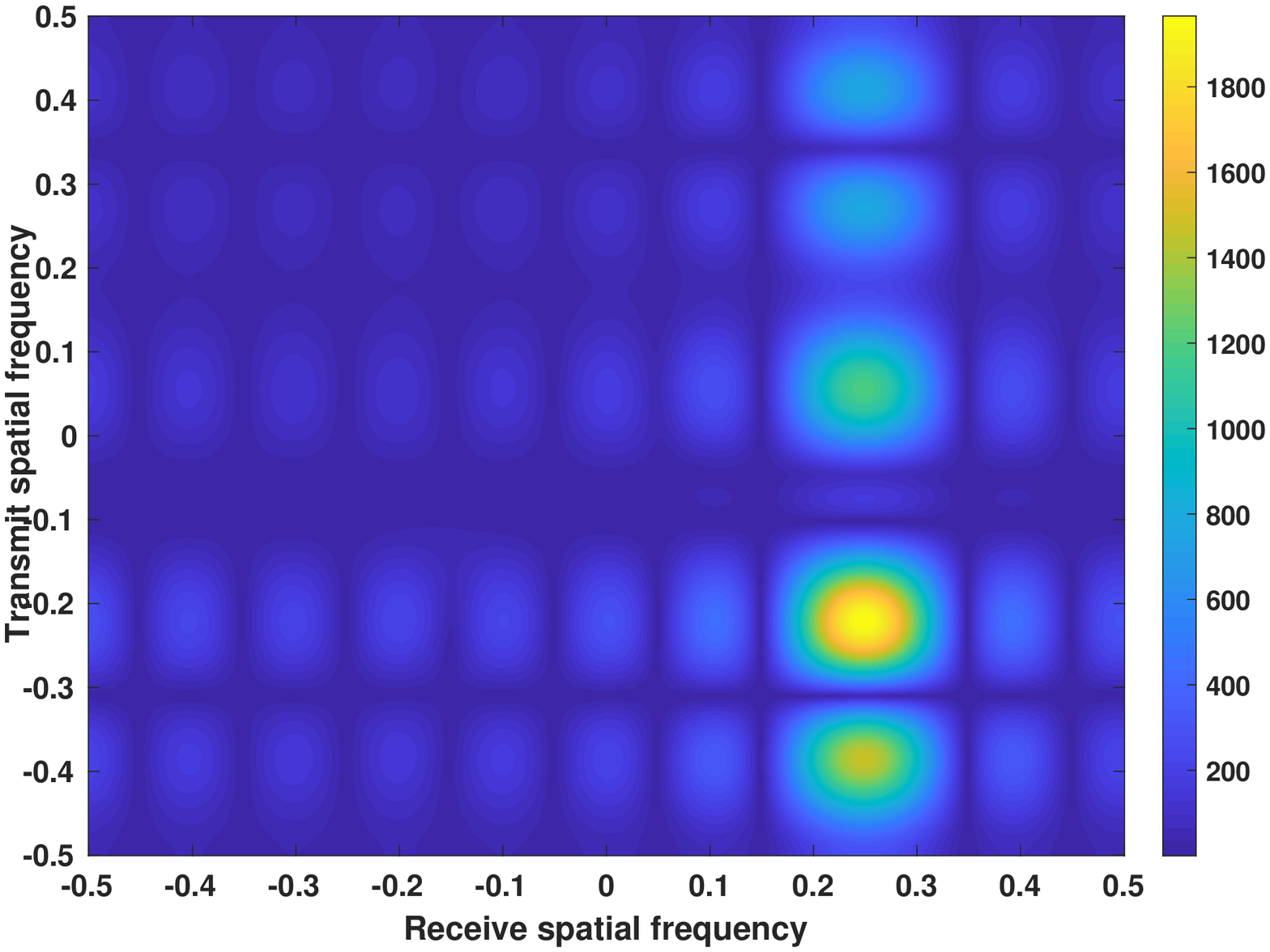}
\end{minipage}
\begin{minipage}{4cm}
\includegraphics[width=4.5cm]{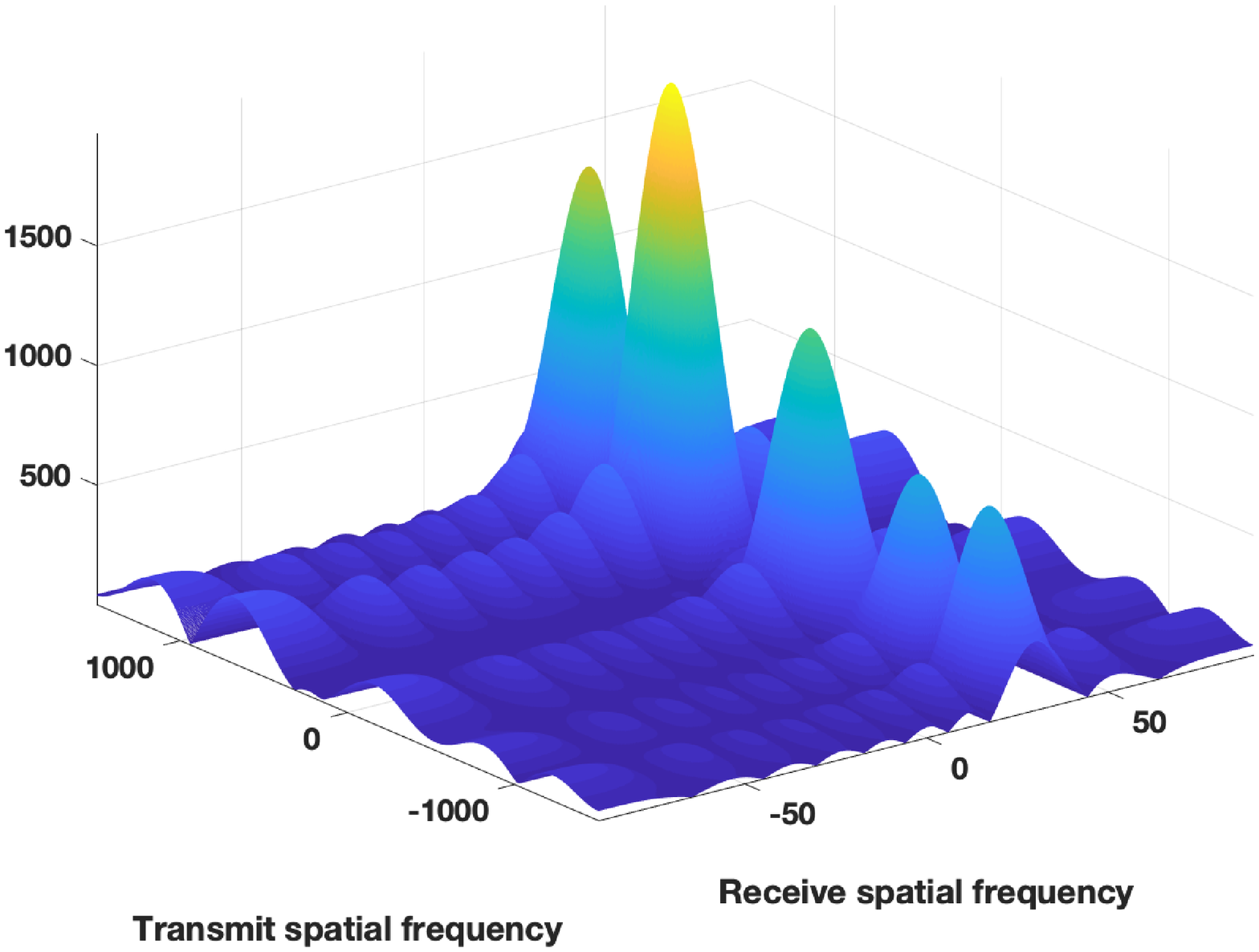}
\end{minipage}
\caption{Pseudo peaks produced by the GoDec method as a solver in mixed jamming signals.}\label{GoDecPeak}
\end{figure}

Low-rank and sparse properties of the components in the mixed matrices have been extensively studied in computational data analysis, with myriad applications ranging from web relevancy data search/analysis to computer vision and image analysis. One of the appealing representatives is the robust PCA \cite{RPCA, RPCA1}:
\begin{align}
\label{robust-PCA}
\min_{\mathbf L, \mathbf S} {\rm{rank}}(\mathbf L) + \lambda||\mathbf S||_0, ~~~{\rm s.t.} ~\mathbf Y = \mathbf L + \mathbf S.
\end{align}
where ${\rm rank}(\mathbf L)$ denotes the rank of $\mathbf L$, the cardinality function $||\mathbf S||_0$ (i.e., $\ell_0$-norm) counts the number of nonzeros of $\mathbf S$, and $\lambda$ is the user parameter that keeps the balance between the rank of $\mathbf L$ and sparsity of $\mathbf S$. As for the variant of PCA, robust PCA is widely used to recover a low-rank matrix $\mathbf L$ from corrupted observations $\mathbf Y$, where corrupted errors $\mathbf S$ are unknown and can be arbitrarily large, but are assumed to be sparse. However, \eqref{robust-PCA} is NP-hard in general as the rank is discrete and nonconvex. A popular and practical solution is to relax \eqref{robust-PCA}, replacing the $\ell_0$-norm with the $\ell_1$-norm, and the rank with the nuclear norm, yielding the following convex surrogate:
\begin{align}
\label{relax-robust-PCA}
\min_{\mathbf L, \mathbf S} ||\mathbf L||_* + \lambda||\mathbf S||_1, ~~~{\rm s.t.} ~\mathbf Y = \mathbf L + \mathbf S,
\end{align}
where the nuclear norm $||\mathbf L||_*$ equals the sum of singular values of $\mathbf L$ and the $\ell_1$-norm is $||\mathbf S||_1 = \sum_{ij}|\mathbf S_{ij}|$. It has been proved that the low-rank and sparse matrices can be recovered under the condition of a unique and precise "low-rank + sparse" decomposition. Even when the decomposition pair is neither low-rank nor sparse, its low-rank and sparse structures can be explored by either low-rank matrix approximation or decomposition \cite{Low-Rank-Approximation}, such as the GoDec method. Although robust PCA \cite{RPCA} and GoDec \cite{GoDec} work well in the case of "low-rank + sparse" decomposition, the corresponding results are not applicable to our problem directly. The reason is that we study the decomposition of $\mathbf Y = \mathbf {X_s} + \mathbf {X_i} + \mathbf {X_e} + \mathbf {X_n}$, where both $\mathbf {X_s}$ and $\mathbf {X_i}$ are low-rank matrices and $\mathbf {X_e}$ is sparse matrix, intrinsically different from robust PCA and GoDec that assumes $\mathbf Y = \mathbf {X_s} + \mathbf {X_e}$ and $\mathbf Y = \mathbf {X_s} + \mathbf {X_e} + \mathbf {X_n}$, respectively. Moreover, if the user parameters change a little, it will affect the estimation performance of robust PCA a lot. Thus,  they all focused on the sparse recovery where all of the entries are real values in the previous methods, while we aim to solve the joint range-and-angle estimation problem for the measured radar data with complex values. The GoDec method is proposed to solve "low-rank + sparse" decomposition problem by bilateral random projections (BRPs)\footnote{Actually, similar with SVD, BRPs is a rank-revealing factorization, which can be used to determine the rank of a matrix. It is easy to verify the rank-revealing property. The interested reader is referred to \cite{Rank-revealing}.} \cite{BRP} based low-rank approximation, where the computational complexity of BRPs-based approximation is lower than that of singular value decomposition (SVD)-based approximation. Nevertheless, the joint range-and-angle estimation problem in our work is "low-rank + low-rank + sparse" decomposition problem. Therefore, the GoDec method cannot be applied straightforwardly, which is verified by simulation result in Fig. \ref{GoDecPeak} where there are some pseudo peaks.

Regarding the data matrix $\mathbf Y $, as for the low-rank properties of $\mathbf {X_s} $ and $\mathbf {X_i} $ and the sparsity property of $\mathbf {X_e} $, $\mathbf Y $ can be decomposed into three independent matrices $\mathbf{L_s}, \mathbf{L_i}$ and $\mathbf{S_e}$, where $\bf L_s$ is the estimate of $\bf X_s$. The task of our solution is to find the matrix $\mathbf {L_s}$ from the corrupted measurement $\mathbf Y $. When the information is essentially redundant, i.e., $ \mathbf {X_s}$ and $ \mathbf {X_i} $ are low-rank and the nonzero elements in $ \mathbf {X_e} $ are sparse. For the specific circumstance of FDA-MIMO model, borrowing the idea of the GoDec method, the approximated decomposition problem stated in \eqref{matrixmodel} is formulated by minimizing the decomposition error with Frobenius norm:
\begin{align}
\label{main-objective}
\min_{\mathbf {X_s}, \mathbf {X_i}, \mathbf {X_j}} &||\mathbf Y - \mathbf {L_s} - \mathbf {S_e} - \mathbf {L_i}||_F^2, \notag\\
{\rm s.t.} ~~~&~{\rm{rank}}(\mathbf {L_s}) \le r_s, \notag\\
~~~&~{\rm{rank}}(\mathbf {L_i}) \le r_i, \notag\\
~~~&~ {\rm card}(\mathbf {S_e}) \le k.
\end{align}
where $r_s$, $r_i$ and $k$ represent the rank of matrices $\mathbf {L_s}$, $\mathbf {L_i}$, and the cardinality of the matrix $\mathbf {S_e}$, respectively. The optimization problem in \eqref{main-objective} restrains the power of noise in the objective function and gives the conditions of the useful signal with the preset rank and cardinality. It is a little different with the traditional robust PCA problem in \eqref{robust-PCA}. The preset ranks $r_s$ and $r_i$, and cardinality $k$ corresponding to the number of sources, hostile interferences and the sparsity of the friendly interferences, respectively. Therefore, the preset value of  $r_s$, $r_i$ and  $k$ require $a$ $priori$ information, which inspires the proposed method in the following section \footnote{It is worthy noting that $r_s$ in \eqref{main-objective} should be the number of targets, which is known $a$ $priori$ in the previous radar task. $r_i$ and $k $ are the number of interferences and nonzero elements in $\mathbf{S_e}$, respectively. In this work, $r_s$ and $r_i$ are selected as 2 and 1, respectively.}.

\subsection{Proposed Range-Angle Estimation Method}

Nonetheless, due to the combinatorial nature of the rank and the cardinality functions, the minimization problem in \eqref{main-objective} is nonconvex and nonsmooth, which is intractable to tackle. To be specific, it is hard to separate low-rank matrix $\mathbf{L_s}$ from noisy measurement $\mathbf Y$, which is also not covered by current models. Motivated by the recent results in  \cite{MMC}, a sampling scheme is added into the procedure of down converting, matched filtering and storing. Therefore, \eqref{matrixmodel} is converted to: 
\begin{align}
\label{Usample}
\mathbf Y_{\mathbf \Xi} = \mathcal{A}^1(\mathbf {L_s}) + \mathcal{A}^2(\mathbf {L_i}) + \mathbf {S_e} + \mathbf {X_n},
\end{align}
where $\mathcal{A}$ is the measurement operator. The $(i, j)$ entry of  $\mathcal{A}(\mathbf {X}) $, denoted by $[\mathcal{A}(\mathbf {X}) ]_{ij}$, can be written as:
\begin{align}
[\mathcal{A}^k(\mathbf {X}) ]_{ij} = \begin{cases}  \mathbf {X}_{ij}, & \mbox{if } (i, j) \in {\mathbf \Xi}, k = 1, 2; \\ 0, & \mbox{otherwise}. \end{cases}
\end{align}
where $\mathbf \Xi \in \{0, 1\}^{MN \times T_\ell}$ indicates the set of observed entries. Let $\mathbf \Xi^1, \mathbf \Xi^2 \in \{0, 1\}^{MN \times T_\ell}$ denote disjoint sets of observed entries, and $\mathbf \Xi := \sum_{k=1}^2\mathbf \Xi^k$. That is to say, each observed entry of $\mathbf Y_{\mathbf \Xi} $ corresponds to an entry in either $\mathbf {L_s}$ or $\mathbf {L_i}$. In words, $\mathbf Y_{\mathbf \Xi} $ contains a mixture of entries from several low-rank matrices. The goal of our work is to recover $\mathbf {L_s}$ in the observed data matrix $\mathbf Y_{\mathbf \Xi} $, which is equivalent to the mixture matrix completion.

Is it possible to separate the low-rank matrix $\mathbf{L_s}$ from $\mathbf Y_{\mathbf \Xi} $? The existing theoretical results in \cite{MMC} give a positive answer. 

{\it Lemma 1:} Without loss of generality, $\mathbf \Xi^k$, $\mathbf {L_s}$ and $\mathbf {L_i}$ satisfy the following assumptions:

1) Each column of $\mathbf \Xi^k$ has either $0$ or $r+1$ non-zero entries.

2) $\mathbf {L_s}$ and $\mathbf {L_i}$ are drawn independently according to an absolutely continuous distribution with respect to the Lebesgue measure on the determinantal variety.

Lemma 1 gives a deterministic condition on $\mathbf \Xi$ to guarantee that $\mathbf {L_s}$ and $\mathbf {L_i}$ can be identified from  $\mathbf Y_{\mathbf \Xi} $. 
The reasonability of \eqref{Usample} is analyzed as follows. As the two low-rank matrices differ sufficiently in both subspaces and singular eigenvalue, they should be modeled in two different low-rank matrices. In practice, the interference matrix might be high in power and its subspace might be spread in transmit spatial domain. In contrast, the target matrix corresponds to a small power and its rank is one considering a single target in the range bin under test. The problem can also be applied for multi-target case under the condition that the number of targets is known $a$ $priori$. For example, in tracking stage of radar system, the target number parameters can be obtained in previous searching stage and can be taken as prior knowledge in current stage. In addition, similar models can be found in one-bit quantization problems \cite{7478040, 8700277, 8788638}.

In order to solve the "low-rank + low-rank + sparse" decomposition problem, we propose a two-step GoDec method for joint range-and-angle estimation with the alternating minimization strategy to avoid SVD and then decompose \eqref{Usample} into two subproblems \cite{YH2019}. For the sake of simplicity, let $\underline{ \mathbf {L_s}} $ and $\underline{ \mathbf {L_i}} $ denote $ \mathcal{A}^1(\mathbf {L_s})$ and  $ \mathcal{A}^2(\mathbf {L_i})$, respectively. We have:
\begin{align}
\label{subproblem1}
\min_{\underline{ \mathbf {L_i}} , \mathbf {S_e}} &||(\mathbf Y_{\mathbf \Xi} - \underline{ \mathbf {L_s}}  - \underline{ \mathbf {L_i}}  - \mathbf {S_e}||_F^2, \notag\\
{\rm s.t.}
~&~{\rm{rank}}(\mathbf {L_i}) \le r_i, \notag\\
~&~ {\rm card}(\mathbf {S_e}) \le k,
\end{align}
and
\begin{align}
\label{subproblem2}
\min_{\underline{ \mathbf {L_s}} } &||(\mathbf Y_{\mathbf \Xi} - \underline{ \mathbf {L_i}} - \mathbf {S_e}) -  \underline{ \mathbf {L_s}}  ||_F^2, \notag\\
{\rm s.t.} ~&~{\rm{rank}}(\mathbf {L_s}) \le r_s,
\end{align}
Regarding the subproblem \eqref{subproblem1}, it is highly nonconvex with respect to $\underline{ \mathbf {L_i}} $ and $\mathbf {S_e}$. Fortunately, based on the framework of low-rank matrix approximation in \eqref{subproblem1}, a fast method as so called GoDec \cite{GoDec}, is proposed to avoid SVD in robust PCA by BRPs. The GoDec method produces approximate decomposition of a general matrix $(\mathbf {Y_\Xi}-\underline{ \mathbf {L_s}} )$ whose exact robust PCA decomposition does not exist due to the additive noise and pre-defined ${\rm rank}(\mathbf{L_i})$ and ${\rm card}(\mathbf{L_s})$. Borrowing the idea of BRPs,  \eqref{subproblem1} is solved by alternative minimization as follows:
\begin{align}
\mathbf {L_i}^{t+1} &= \arg\min_{{\rm rank}(\mathbf{L_i})\le r_i}||(\mathbf {Y_\Xi} -\underline{ \mathbf {L_s}} ) - \mathbf {S_e}^t - \underline{ \mathbf {L_i}} ||_F^2\notag\\
&= (\mathbf {Y_\Xi}- \underline{ \mathbf {L_s}} ) \mathbf{A}_1 (\mathbf A_2^H (\mathbf{Y_\Xi} - \underline{ \mathbf {L_s}} ) \mathbf{A}_1)^{-1} ((\mathbf{Y_\Xi} - \underline{ \mathbf {L_s}} )^H\mathbf A_2)^H\label{L-update}\\
\mathbf {S_e}^{t+1} &= \arg\min_{{\rm card}(\mathbf{S_e})\le k}||(\mathbf{Y_\Xi} - \underline{ \mathbf {L_s}} ) - \mathbf {S_e} - \underline{ \mathbf {L_i}}^{t+1}||_F^2\notag\\
&= \mathcal P_\Omega(\mathbf {Y_\Xi} - \underline{ \mathbf {L_i}}^{t+1}),\label{Xi-solution}
\end{align}
where $\mathbf A_2 =(\mathbf {Y_\Xi}- \underline{ \mathbf {L_s}} ) \mathbf{A}_1$ and $\mathbf A_1 = (\mathbf {Y_\Xi} -\underline{ \mathbf {L_s}} )^H\mathbf A_2$ are updated to improve the approximation precision of $\mathbf{L_i}^{t+1}$. $\mathbf{S_e}^{t+1}$ is updated via entry-wise hard thresholding of $(\mathbf Y -\mathbf{L_s}- \mathbf{L_i}^{t+1})$. $\Omega: |(\mathbf Y - \mathbf {L_i}^{t+1})_{i,j}\in\Omega|\ge|(\mathbf Y - \mathbf {L_i}^{t+1})_{i,j}\in{\bar \Omega}|, |\Omega|\le k$ and $\mathcal P_\Omega(\mathbf Y)$ is the projection to extract the first $k$ largest entries of $|\mathbf Y - \underline{ \mathbf {L_s}} -\underline{ \mathbf {L_i}}^{t+1}|$, where the subscript $(\cdot)_{i,j}$ denotes the element located at the $i$th row and $j$th column of the matrix $\mathbf Y$. In this work, the GoDec method is initialized with $\mathbf {L_s} = \mathbf 0$, and $\mathbf 0$ denotes a matrix with all zeros. The detected results in \eqref{Xi-solution} may be polluted by the noise term. The noise term may disturb the detection and result in some false targets, which is demonstrated from Fig. \ref{GoDecPeak}. Also, the cardinality $k$ preset in the optimization problem determines the detected number of nonzero points in the sparse matrix. Herein, we propose two-step GoDec with $a$ $priori$ $r_s$ in \eqref{subproblem2} to delete the pseudo peaks in Fig. \ref{GoDecPeak}. 

Next, for the Frobenius norm minimization in \eqref{subproblem2} with rank constraint, it is easily handled as it is a least squares problem. To guarantee the solution \eqref{Xi-solution} to perform good performance, especially for singular values of $(\mathbf {Y_\Xi} -\underline{\mathbf{L_s}})$ decaying slowly, $(\mathbf {Y_\Xi} -\underline{\mathbf{L_s}})$ is modified as $((\mathbf {Y_\Xi} -\underline{\mathbf{L_s}})(\mathbf{Y_\Xi} -\underline{\mathbf{L_s}})^H)^q(\mathbf{Y_\Xi} -\underline{\mathbf{L_s}}) (q \ge 0)$  based on the power scheme \cite{PS}. Both $(\mathbf {Y_\Xi} -\underline{\mathbf{L_s}})$ and $((\mathbf {Y_\Xi} -\underline{\mathbf{L_s}})(\mathbf {Y_\Xi} -\underline{\mathbf{L_s}})^H)^q(\mathbf {Y_\Xi} -\underline{\mathbf{L_s}})$ share the same singular vectors, but the singular values of the latter decay faster than $(\mathbf {Y_\Xi} -\underline{\mathbf{L_s}})$. Therefore, the solution of \eqref{subproblem2} is obtained by
\begin{align}\label{Xs-update1}
\mathbf{L_s}^{t+1}&= \mathbf{\bar Y} \mathbf{\bar A}_1 (\mathbf{\bar A}_2^H \mathbf{\bar Y}  \mathbf{\bar A}_1)^{-1} (\mathbf{\bar Y}^H\mathbf{\bar A}_2)^H, 
\end{align}
where $\mathbf{\bar Y}=\mathbf {Y
_\Xi}-\underline{\mathbf{L_i}}^{t+1}-\mathbf{S_e}^{t+1}$, $\mathbf{\bar A}_2 =\mathbf{\bar Y} \mathbf{\bar A}_1$, and $\mathbf{\bar A}_1 = \mathbf{\bar Y}^H\mathbf{\bar A}_2$. The BRPs is employed instead of SVD for low-rank approximation in \eqref{Xs-update1} to accelerate the convergence and reduce the computational burden. In order to obtain the approximation of $\mathbf{L_s}$ with rank $r_s$, we calculate the QR decomposition of $\mathbf{\bar A}_1 $ and $\mathbf{\bar A}_2$, that is:
\begin{align}
\mathbf{\bar A}_1 = \mathbf Q_1\mathbf R_1, ~~\mathbf{\bar A}_2 = \mathbf Q_2\mathbf R_2. 
\end{align}  
Then a fast rank-$r$ approximation of  $\mathbf{L_s}$ can be updated as:
\begin{align}\label{Xs-update}
\mathbf{L_s}^{t+1}&= \mathbf{Q}_2[\mathbf R_2 (\mathbf{\bar A}_2^H \mathbf{\bar Y}  \mathbf{\bar A}_1)^{-1} \mathbf{R}_1^H]^{1\over{2q+1}}(\mathbf{Q}_1)^H, 
\end{align}
The power scheme modification in \eqref{Xs-update} requires the inverse of an $r\times r$ matrix, and QR decomposition of two projection matrices and five matrix multiplications. If we do not perform the power scheme modification, i.e., $q = 0$, then the rank-$r$ approximation is obtained by \eqref{Xs-update1}. The complexity of GoDec method to deal with \eqref{subproblem1} is $\mathcal O(r_s^3 + 2T_\ell r_s^2 +4MNT_\ell)$ flops to update $\bf X_j$ and $\bf X_i$ per iteration, $\mathcal O(r_s^3 + 2T_\ell r_s^2 + MNT_\ell + 4qMNr_s + 5MNr_s)$ is required to update $\bf X_s$ using one BRPs approximation. The BRPs operation needs much less flops per iteration compared with the SVD operation.  Above all, the proposed two-step GoDec method is summarized in Algorithm 1. 
\begin{algorithm}[http]
\caption{Two-step GoDec method via alternating minimization}
\label{Two-step GoDec method}
\begin{algorithmic}
\REQUIRE {$\mathbf Y, M, N, T_\ell$}\\
\STATE{{\bf Initialize:} $r_s=2, r_i=1, k $}
\FOR{$t = 0, 1, \cdots$}
\REPEAT
\STATE{Update $\mathbf {L_i}^{t+1}$ according to  \eqref{L-update};}
\STATE{Update $\mathbf {S_e}^{t+1}$ according to \eqref{Xi-solution};}
\STATE{Update $\mathbf {L_s}^{t+1}$ according to \eqref{Xs-update};}
\UNTIL{convergence}
\ENDFOR
\ENSURE{$\mathbf {L_s}$}\\ 
\end{algorithmic}
\end{algorithm}

\begin{figure}[!htbp]
\begin{center}
\includegraphics[width=8.5cm]{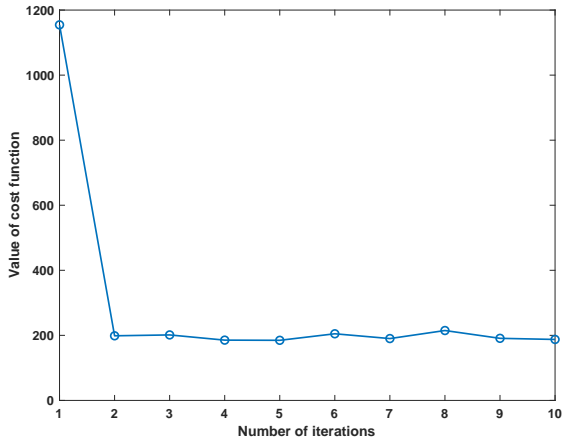}
\end{center}
\caption{Convergence behavior}
\label{Convergence}
\end{figure}

The convergence of  the GoDec method has been proved in \cite{GoDec}, but it makes the theoretical proof of the convergence of the proposed two-step GoDec method challenging due to the nonconvexity of the rank minimization between alternating minimization. Although the convergence is not proved theoretically, we observe that the proposed method always converges in the simulations. Thus, it is deemed that the proposed two-step GoDec method is empirically convergent in practice, where the convergence rate of the proposed method is plotted in Fig. \ref{Convergence}, where Algorithm 1 converges to the true solution in one iteration in the mixed jamming signals.

Additionally, the performance of traditional robust PCA method will dramatically deteriorate for strong power of noise, which can be explained in \eqref{main-objective} because the objective function is actually equal to the energy of noise. The GoDec method is the representative of the traditional robust PCA method. If the noise level is low, the estimates of $\mathbf{L_i}$, $\mathbf{S_e}$ and $\mathbf{L_s}$ are quite accurate. However, the estimated errors will increase in high noise level. For the proposed method, it has more robust performance compared with the traditional robust PCA methods, since the rank minimization in  \eqref{subproblem2} helps to suppress the false targets resulting from the noise term.

In the sequel, with the estimated solution of $\mathbf {L_s}$, its singular value decomposition can be expressed as:
\begin{align}
  \mathbf{L_s} = \mathbf{U \Lambda V}
\end{align}
where the columns of $\mathbf U$ and $\mathbf V$ are the left and right singular vectors, respectively, and $\mathbf \Lambda$ is a diagonal matrix whose diagonal elements are the singular values. The range bin indexes of targets can be obtained as $\mathcal{I} = \{i | \mathbf{V}\{ \cdot , i\} > 0 \}$, which is due to the distribution of target in range. Given a particular range bin index $i$, the target signal vector is constructed as:
\begin{align}
  \mathbf{l_s}^i = \mathbf{L_s}(\cdot , i).
\end{align}
Generally, if there is only one possible target within the particular range bin, the range and angle parameters of target can be estimated by 
\begin{align}
  \{r,\theta \} = \arg_{r,\theta} \max \textsl{FFT}_{2D} \{ \mathbf{Z_s}^i \},
\end{align}
where $\mathbf{Z_s}^i = \text{mat} \{ \mathbf{l_s}^i \}$ is the signal matrix corresponding to the considered range bin, $\text{mat}\{\cdot\}$ is to transform a vector into a matrix, and $\textsl{FFT}_{2D}$ denotes the 2-D Fourier transform.  If multiple targets appear in the same range bin, it requires to estimate the number of independent targets first. The CLEAN approach \cite{CLEAN} can be applied in the multiple-target parameter estimation. This is also important to understand the model where two low-rank matrices are isolated. Besides, sparse recovery-based methods can be used to obtain the joint range and angle parameters of targets, where the minimization problem is expressed as:
\begin{align}
  \{r,\theta \} = \arg_{r,\theta} \min{ \{ \| \mathbf{l_s}^i - \mathbf{D}(r,\theta) \mathbf{a} \|_2 + \| \mathbf{a} \|_0 \} },
\end{align}
in which $\mathbf{D}(r,\theta)$ is the redundant dictionary in the two-dimensional Fourier transform dimension. As it shows, $\mathbf{D}(r,\theta)$ associates the range and angle parameters, forming the grids in the Fourier transform dimension. Finally, it is possible to get the range and angle estimation with the sparse recovery result.

\section{SIMULATION RESULTS}
In this section, the complex radar data is presented to demonstrate the effectiveness of the proposed method, where the specific parameters of the FDA-MIMO radar are shown in Table \ref{Table_param}.
All the simulation results are conducted using a computer with a 3.4 GHz Intel core i7 CPU and 4 GB RAM, under a 64-bit Microsoft Windows 7 operating system.

\begin{table}
\caption{PARAMETERS OF FDA-MIMO RADAR}
\begin{tabular*}{8.8cm}{lll}
\hline
\hline
Parameter & Value   \\
\hline
Reference frequency  & 10GHz  \\
Frequency increment  & 301250Hz \\
Waveform bandwidth & 15MHz\\
Interference angle & $0^\circ$ \\
INR & 30dB \\
Element number & 6 \\
Element spacing & 0.015m \\
Number of pulses & 100 \\
\hline
\hline
\end{tabular*}\label{Table_param}
\end{table}

\begin{figure}[!htbp]
\centering
\begin{minipage}{4cm}
\includegraphics[width=4.5cm]{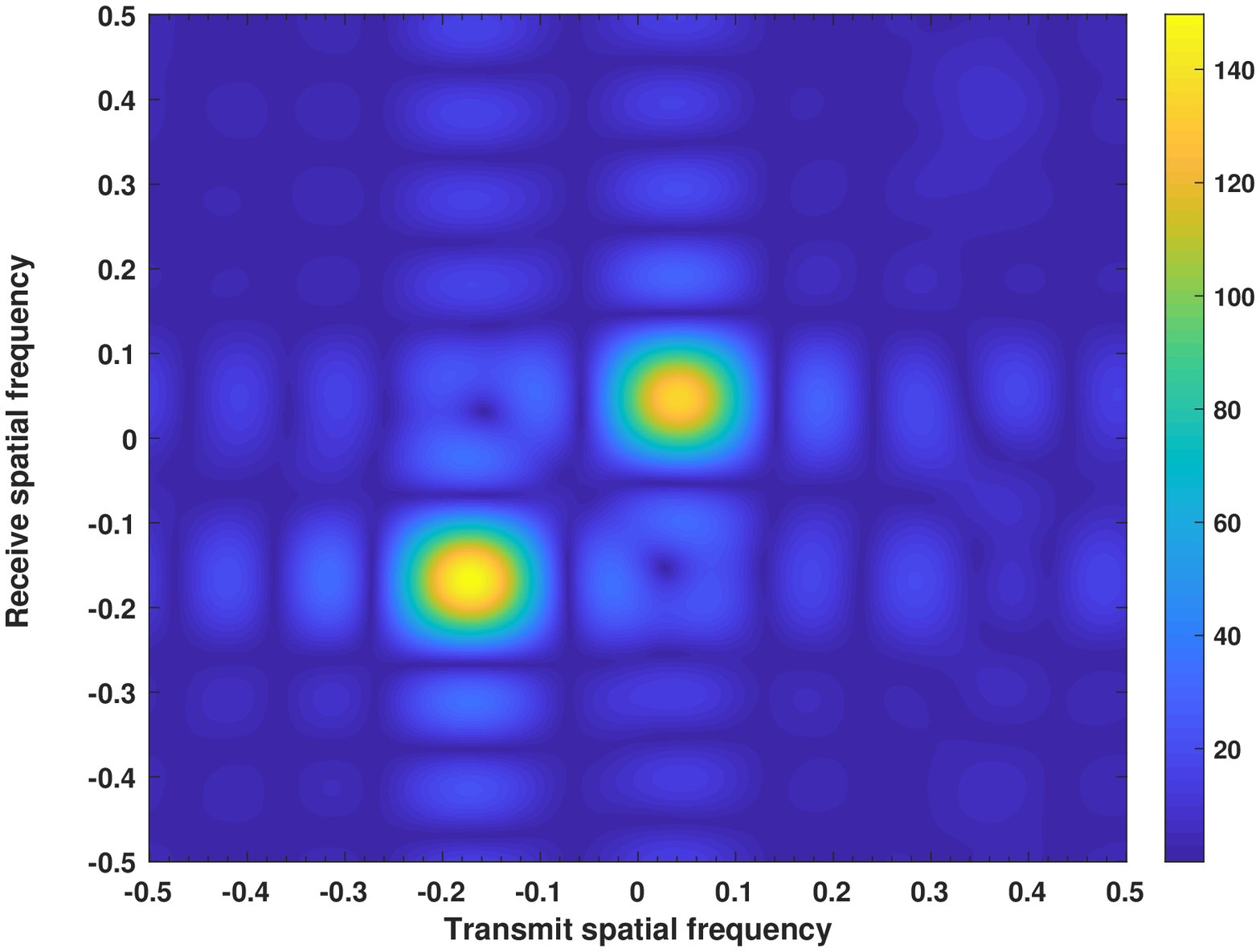}
\centerline{(a) }
\end{minipage}
\begin{minipage}{4cm}
\includegraphics[width=4.5cm]{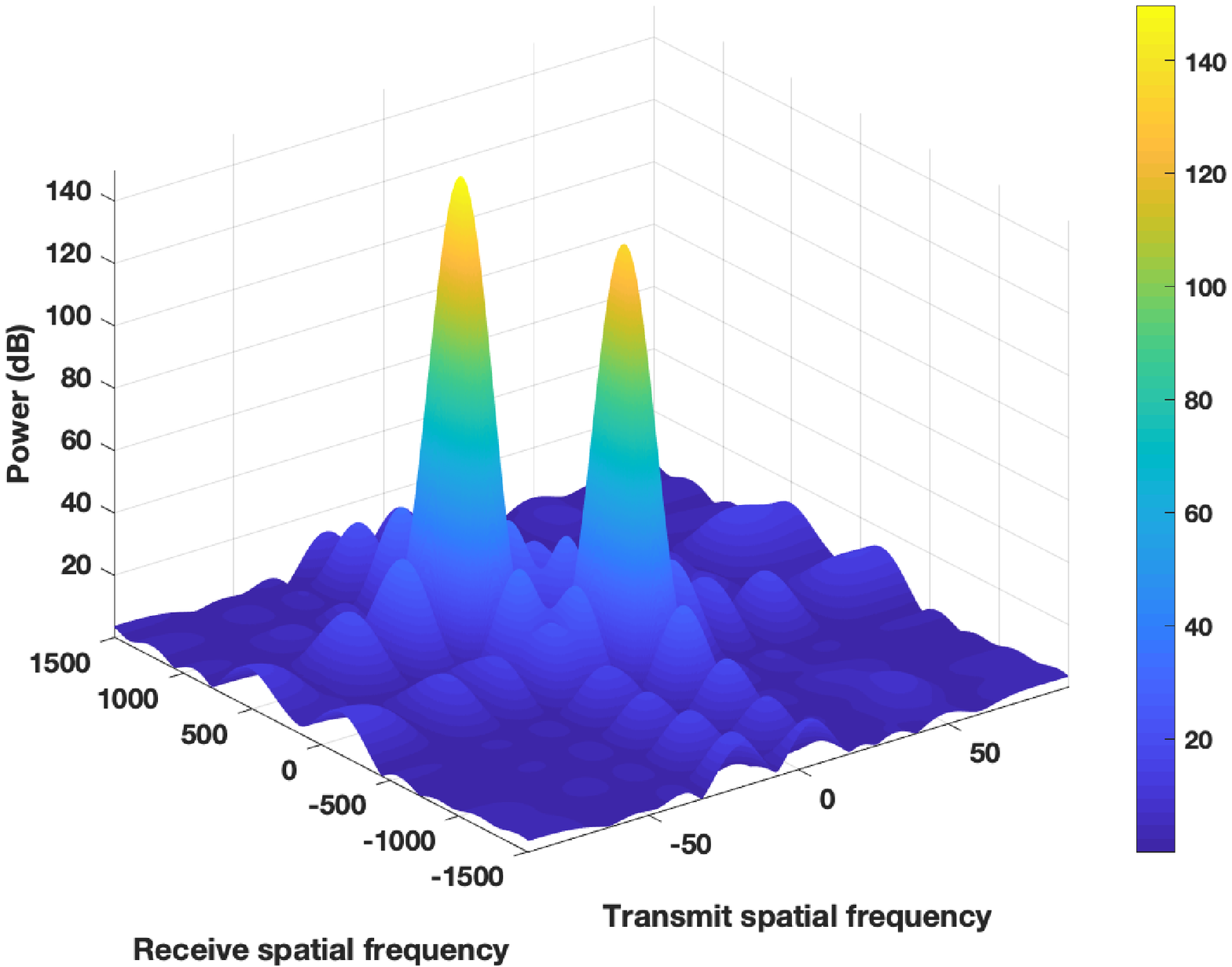}
\centerline{(b) }
\end{minipage}
\caption{(a) Spectrum distribution of two different targets in joint transmit-receive spatial frequency domain. (b) Spectrum distribution in three-dimension view.}\label{S1}
\end{figure}

\begin{figure}[!htbp]
\centering
\begin{minipage}{4cm}
\includegraphics[width=4.5cm]{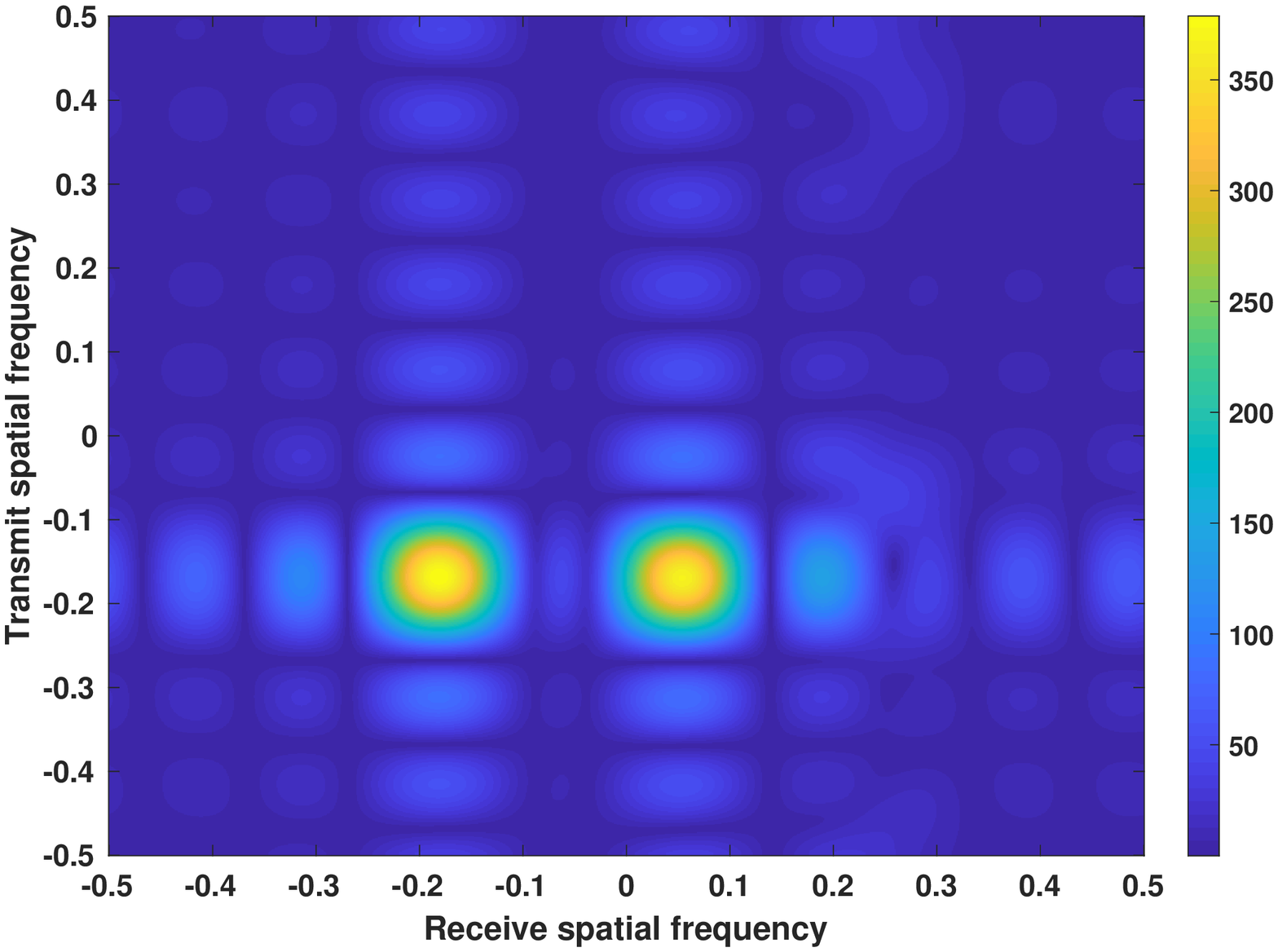}
\centerline{(a) }
\end{minipage}
\begin{minipage}{4cm}
\includegraphics[width=4.5cm]{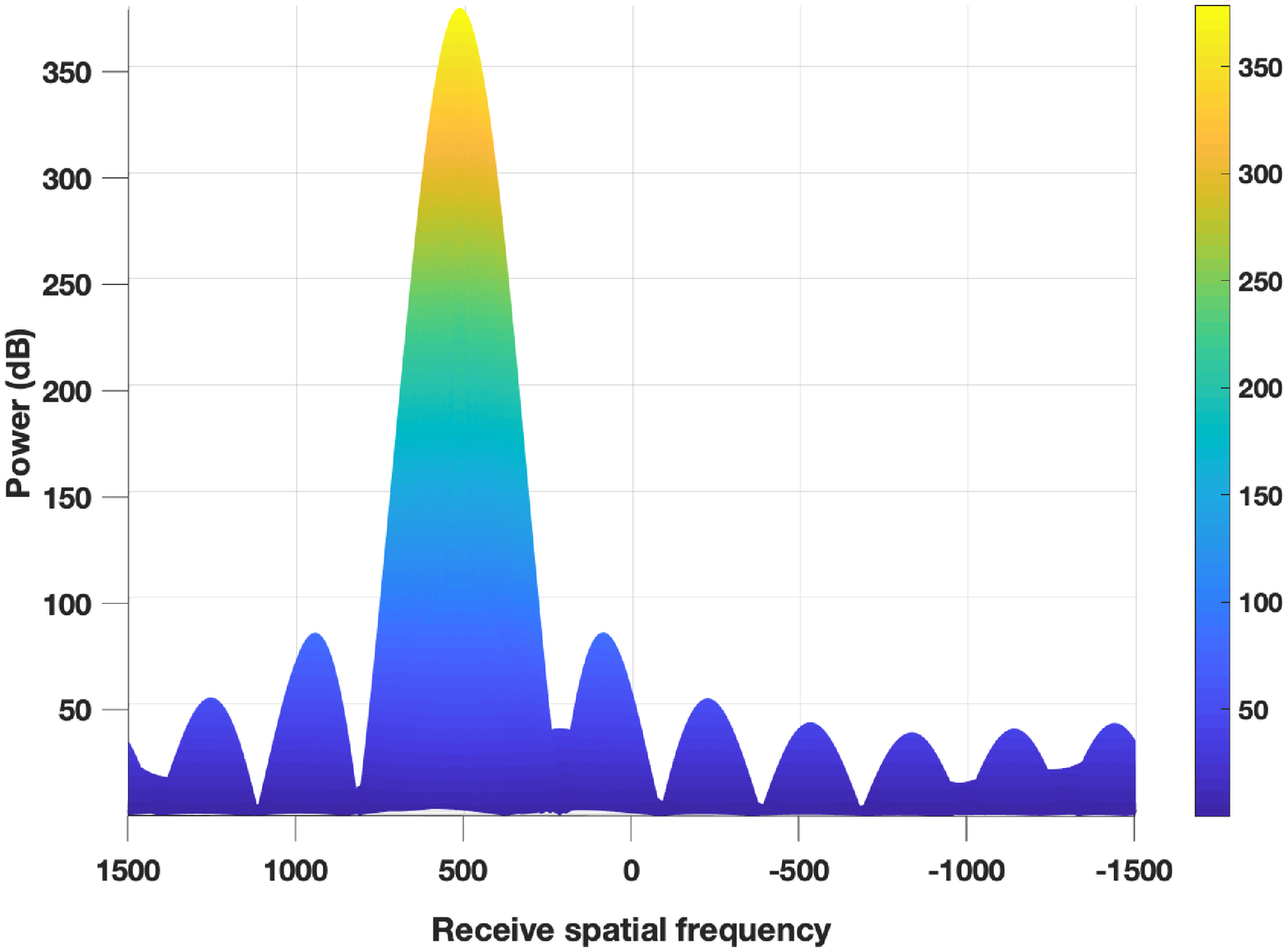}
\centerline{(b) }
\end{minipage}
\caption{(a) Spectrum distribution of two targets with different angles but same ranges in joint transmit-receive spatial frequency domain. (b) Spectrum distribution in receive spatial frequency domain.}\label{SameRange}
\end{figure}

\begin{figure}[!htbp]
\centering
\begin{minipage}{4cm}
\includegraphics[width=4.5cm]{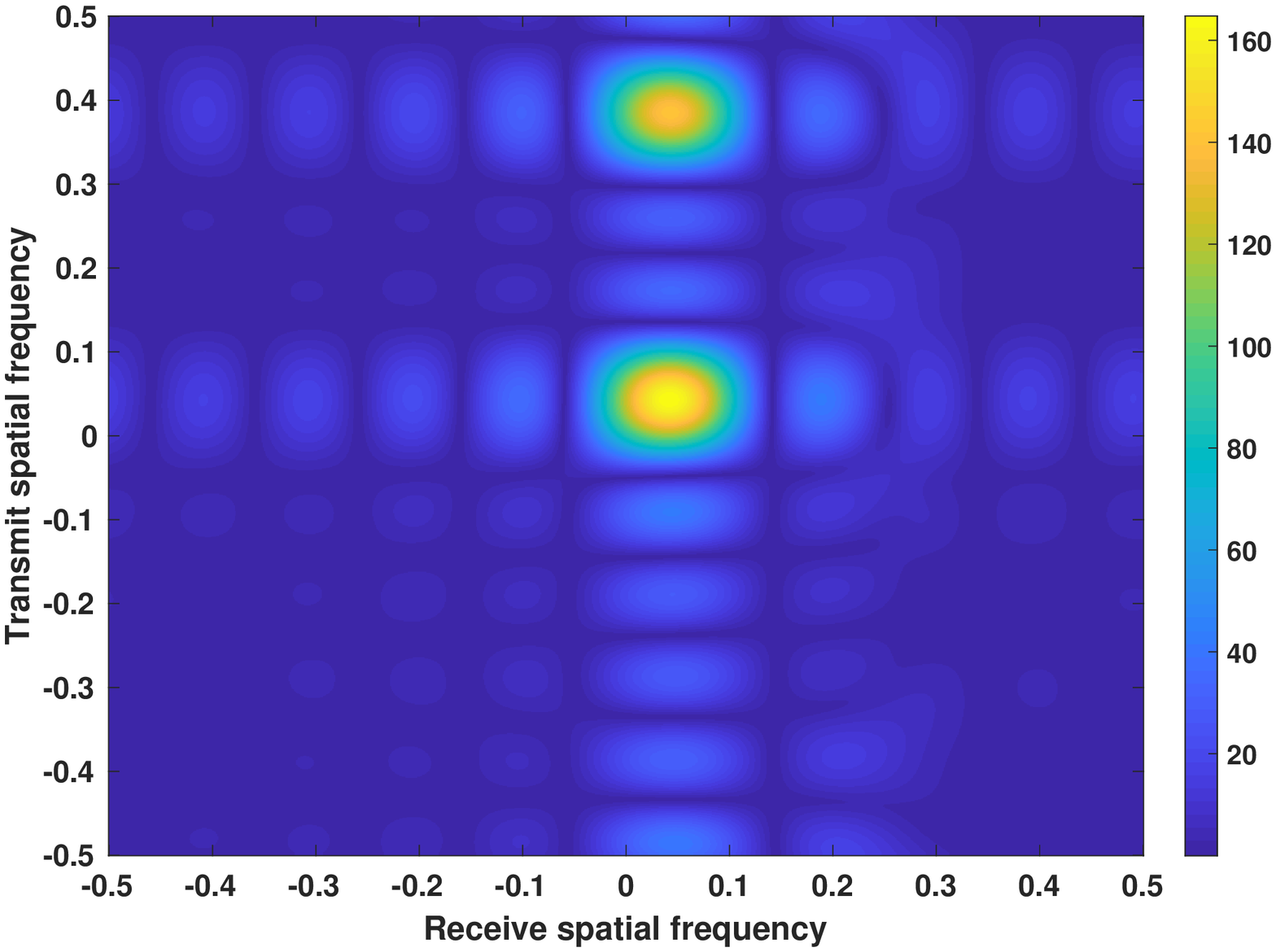}
\centerline{(a) }
\end{minipage}
\begin{minipage}{4cm}
\includegraphics[width=4.5cm]{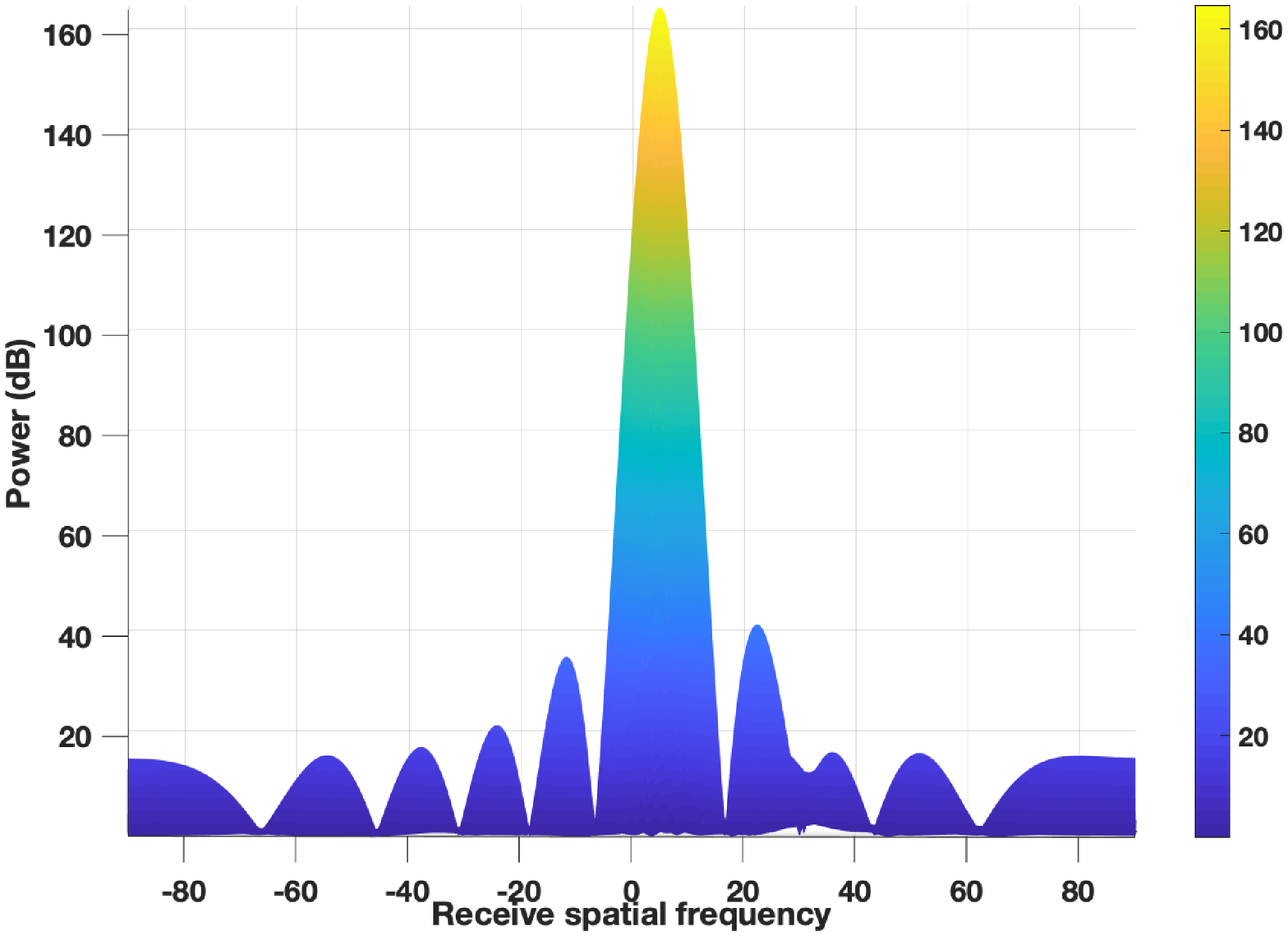}
\centerline{(b) }
\end{minipage}
\caption{(a) Spectrum distribution of two targets with different ranges but same angles in joint transmit-receive spatial frequency domain. (b) Spectrum distribution in transmit spatial frequency domain.}\label{SameAngle}.
\end{figure}

In the first example, we evaluate the spectrum distributions of two targets solved by the proposed method in the range bin under test, shown in Figs. \ref{S1} to \ref{SameAngle}. Fig. \ref{S1} shows the power spectrum of two targets with the locations $(-20^\circ, -5 ~{\rm km})$ and $(5^\circ, 1.5 ~{\rm km})$. We can see that the two peaks corresponding to the two target locations are clearly visible and are not masked by the pseudo peaks, which implies that the proposed method enjoys better resolution performance. We also consider two scenarios: 1) having the same range but different angles, i.e., i.e., $(-20^\circ, 5 ~{\rm km})$ and $(5^\circ, 5 ~{\rm km})$; 2) having the same angle but different ranges, i.e., $(5^\circ, 1.5 ~{\rm km})$ and $(5^\circ, 5 ~{\rm km})$,  as shown in Fig. \ref{SameRange} and Fig. \ref{SameAngle}, respectively. Since two targets have the same ranges in Fig. \ref{SameRange} (b), they cannot be separated in range dimension but it can be easily identified in angle dimension because of the MIMO property, seen from Fig. \ref{SameRange} (a). While the two targets share the same angle with different ranges, they cannot be distinguished in  angle dimension, but can be separated in range domain due to the property of FDA.

\begin{figure}[!htbp]
\begin{center}
\includegraphics[width=8.5cm]{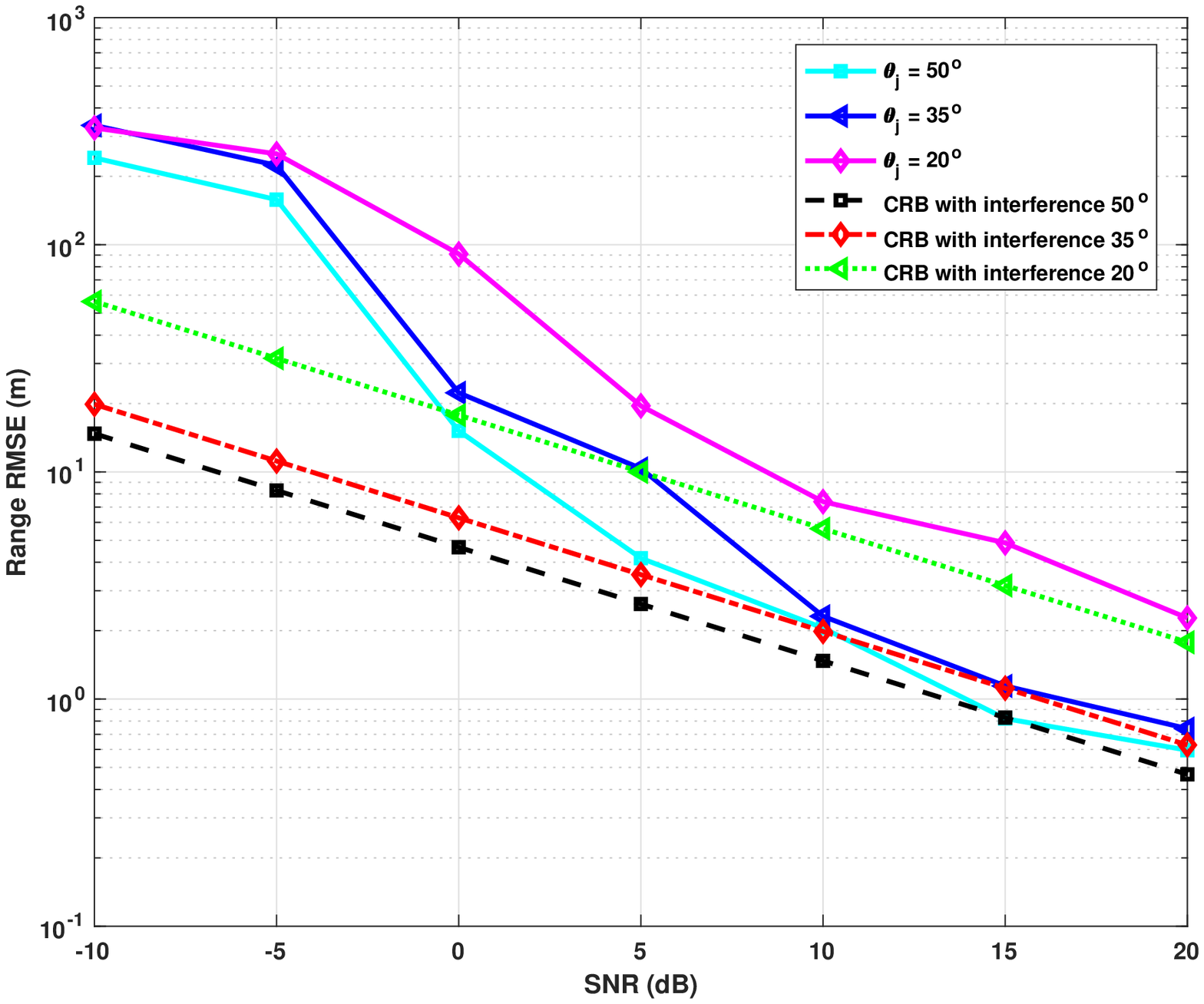}
\end{center}
\caption{RMSE versus SNR for range estimation}
\label{Theta-range}
\end{figure}

\begin{figure}[!htbp]
\begin{center}
\includegraphics[width=8.5cm]{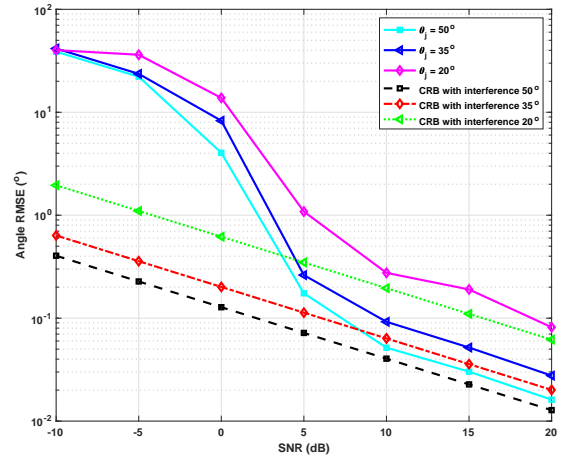}
\end{center}
\caption{ RMSE versus SNR for angle estimation}
\label{Theta-theta}
\end{figure}

In the second example, as shown in Fig. \ref{Theta-range} and Fig. \ref{Theta-theta}, the root mean square error (RMSE) of the range and angle estimates with respect to input signal-to-noise ratio (SNR) are used to test the effect of barrage jamming, respectively, where a target of interest is supposed to reflect a plane-wave that impinges on the array from $\theta = 0^\circ$ and the slant range of $r = 5$ km. We evaluate the effect of the barrage jamming signal from the different directions of $\theta_j = \{20^\circ, 35^\circ, 50^\circ\}$. To this end, 800 Monte Carlo trials have been carried out, where the number of snapshots is 100. It can be seen from the simulation results that when $\theta_j$ is progressively far away from the impinging angle of desired signal, i.e., $\theta = 0^\circ$, the estimation accuracy becomes better, which is also verified with the comparison of Cramer-Rao bound (CRB). 
When the angle of barrage jamming signal changes from $50^\circ$ to $20^\circ$, the CRBs become larger. Specially, when $\theta_j$ is fixed at $0^\circ$, it is observed that the proposed method  fails to estimate the localization parameters of target. However, how to dissolve the case when the angle of barrage jamming is the same as that of the target, is not our goal in this work, which can be treated as future work.



\begin{figure}[!htbp]
\begin{center}
\includegraphics[width=8.5cm]{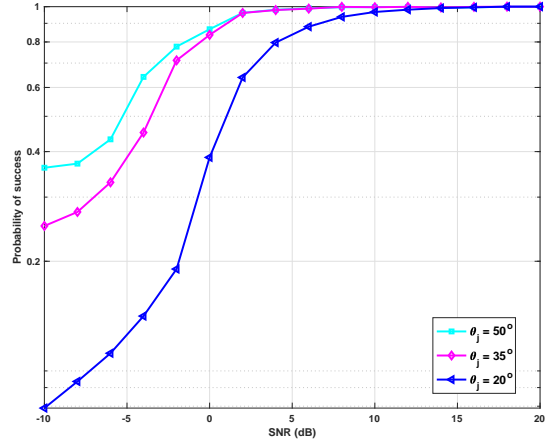}
\end{center}
\caption{Probability of success versus SNR in angle estimation}
\label{theta_prob}
\end{figure}

\begin{figure}[!htbp]
\begin{center}
\includegraphics[width=8.5cm]{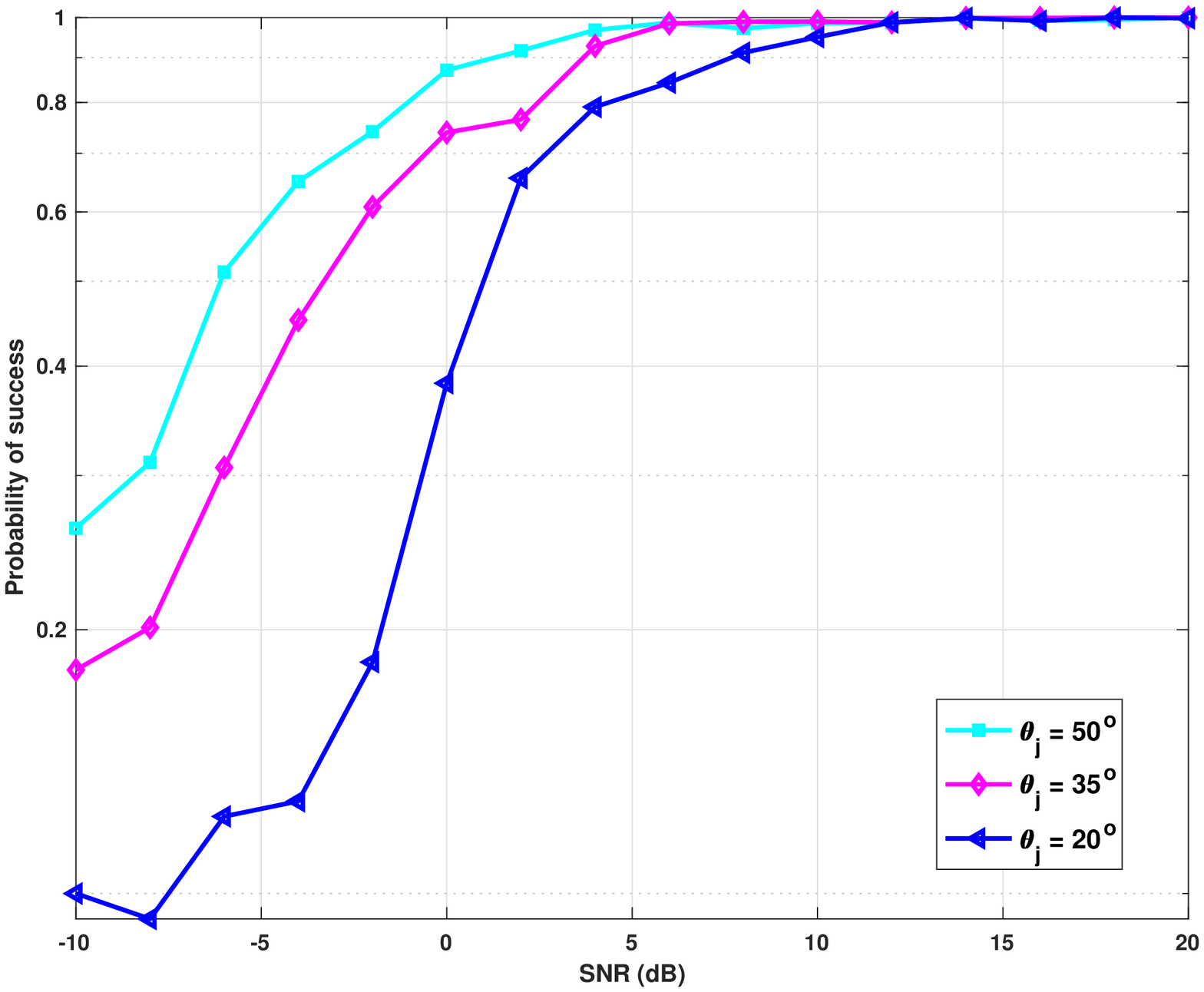}
\end{center}
\caption{Probability of success versus SNR in range estimation}
\label{range_prob}
\end{figure}

To further evaluate the performance of the proposed solution, the probability of success versus SNR in the case of different directions of barrage jamming signals is examined and the results are plotted in Figs. \ref{theta_prob} and \ref{range_prob}. The probability of success is computed as the ratio between the number of successful runs and the total number of the independent runs. A trial is regarded as a successful one when the absolute deviation between the estimated and true parameters, i.e., $|{1\over Q}\sum_{\ell=1}^Q(\hat{\omega}^{(\ell)}-\omega)|$, is less than $ 10^{-2}$ for angle estimation and $10 m$ for range estimation, respectively. Here, $\hat{\omega}^{(\ell)}$ denotes the source parameter estimate obtained by the proposed algorithm at the $\ell$th Monte Carlo run, and $Q$ is the number of Monte Carlo trials. It is concluded that the proposed method with $\theta_j \ge 35^\circ$ exhibits a $100\%$ correct resolution probability when SNR $\ge 5$ dB, where the target is fixed at $\theta = 0^\circ$. We also notice that the proposed scheme fails to detect the target when $\theta_j$ close to the angle of the desired signal. Meanwhile, the performance with $\theta_j = 50^\circ$ is slightly better than that at $\theta_j = 35^\circ$.

\section{CONCLUSIONS}
In this paper, we present a novel joint range and angle estimator for FDA-MIMO radar, where the non-Gaussian burst jamming signal from friendly radar or other working radio equipment is included. Our motivation is twofold, namely, to capture a more general mixed jamming signals removal model in practice, and to suppress the false targets resulting from the mixed jamming signal in the existing GoDec method. Therefore, a two-step GoDec via alternating minimization algorithm is proposed to localize the sources in FDA-MIMO radar, which is parameter-free and easily implemented. Compared with the traditional MIMO radar, FDA-MIMO radar is superior since the targets with same angle but different ranges can be dissolved with the help of FDA. Moreover, with $a$ $priori$ knowledge of the targets, the effectiveness of the proposed scheme on the range and angle estimation is validated by simulation results. Thus, as a future work, we expect that the parameter estimation accuracy can be improved especially when the angle of target is close to that of barrage jamming signal, and an unbiased joint range and angle estimator can be devised for mixed jamming signal removal.

\section*{APPENDIX}
\subsection{Derivation of Equation \eqref{Rf}}
To derive \eqref{Rf}, we have:
\begin{align}
x_n(t) &=\sum_{m = 1}^{M}\sqrt{E\over M}\xi\phi_m(t-\tau_{m,T}-\tau_{n,R})e^{j2\pi f_mt}{e^{j2\pi f_d(t-{2r\over c})}}\notag\\
&~~~~ \times {e^{{-j2\pi f_m}({2r\over c}-{{(m-1)d_T\sin\theta}+{(n-1)d_R\sin\theta}\over c}))}}\notag\\
&~~~~ \times e^{j2\pi f_d{{(m-1)d_T\sin\theta}+{(n-1)d_R\sin\theta}\over c}} \label{app1}\\
&\approx\sum_{m = 1}^{M}\sqrt{E\over M}\xi\phi_m(t-\tau_{m,T}-\tau_{n,R}) e^{j2\pi f_mt} {e^{j2\pi f_d(t-{2r\over c})}} \notag\\
&~~~~ \times {e^{{-j2\pi f_m}({2r\over c}-{{(m-1)d_T\sin\theta}+{(n-1)d_R\sin\theta}\over c}))}} \label{app2}\\
&=\sum_{m = 1}^{M}\sqrt{E\over M}\xi\phi_m(t-\tau_{m,T}-\tau_{n,R})e^{j2\pi f_mt}  \notag\\
&~~~~ \times {e^{j2\pi (f_d(t-{2r\over c})- f_0{2r\over c})}} \notag\\
&~~~~ \times {e^{{-j2\pi }({{(m-1)\Delta f2r \over c}}-{(m-1)d_T f_0\sin\theta\over c}-{(n-1)d_Rf_0\sin\theta\over c})}} \notag\\& ~~~~\times {e^{j2\pi({(m-1)^2\Delta fd_T\sin\theta\over c}+{(m-1)(n-1)\Delta fd_R\sin\theta\over c})}} \label{app3}\\
&\approx\sum_{m = 1}^{M}\sqrt{E\over M}\xi\phi_m(t-\tau_{m,T}-\tau_{n,R}) e^{j2\pi f_mt} \notag\\
&~~~~ \times {e^{j2\pi (f_d(t-{2r\over c})- f_0{2r\over c})}} \notag\\
&~~~~ \times {e^{{-j2\pi }({{(m-1)\Delta f2r \over c}}-{(m-1)d_T f_0\sin\theta\over c}-{(n-1)d_Rf_0\sin\theta\over c})}}\\
&=\sum_{m = 1}^{M}\sqrt{E\over M}\xi\phi_m(t-\tau_{m,T}-\tau_{n,R}) e^{j2\pi ((m-1)\Delta f)t} e^{j2\pi f_0t}\notag\\
&~~~~ \times {e^{j2\pi (f_d(t-{2r\over c})- f_0{2r\over c})}} \notag\\
&~~~~ \times {e^{{-j2\pi }({{(m-1)\Delta f2r \over c}}-{(m-1)d_T f_0\sin\theta\over c}-{(n-1)d_Rf_0\sin\theta\over c})}}
\label{app5}
\end{align}

For the approximation \eqref{app2}, due to $d_T = {c\over{2f_0}}$ and $d_R = {c\over{2f_0}}$, we obtain:
\begin{align}
f_d({{(m-1)d_T\sin\theta}+{(n-1)d_R\sin\theta}\over c}) \notag\\
= {f_d\over2f_0}({{(m-1)\sin\theta}+{(n-1)\sin\theta}}). \notag
\end{align}
Since $f_0 \gg f_d$, this term is negligible for the sake of simplicity. On the other hand,  the last term of \eqref{app3} is small and can be ignored.

\bibliography{IEEEabrv,IEEEabrv.bib}
\bibliographystyle{ieeetr}


\end{document}